\documentclass{article}
\usepackage{spconf,amsmath,graphicx,balance}
\usepackage{amssymb}
\usepackage{xcolor}
\usepackage{hyperref}
\usepackage{url}
\hypersetup{
    colorlinks=true,
    linkcolor=blue,
    filecolor=magenta,      
    urlcolor=cyan,
}
\usepackage{enumitem}
\setlist[1]{itemsep=1pt}

\usepackage{booktabs}
\graphicspath{{figures/}} 

\title{AECMOS: A speech quality assessment metric for echo impairment}
\name{Marju Purin, Sten Sootla, Mateja Sponza, Ando Saabas, Ross Cutler 
\parbox{\linewidth}{\centering}
}
\address{Microsoft Corporation}

\begin{document}
\ninept
\maketitle
\begin{abstract}
Traditionally, the quality of acoustic echo cancellers is evaluated using intrusive speech quality assessment measures such as ERLE \cite{g168} and PESQ \cite{p862}, or by carrying out subjective laboratory tests \cite{itut_p831, itut_p832}. Unfortunately, the former are not well correlated with human subjective measures, while the latter are time and resource consuming to carry out \cite{cutler2020crowdsourcing}. We provide a new tool for speech quality assessment for echo impairment which can be used to evaluate the performance of acoustic echo cancellers. More precisely, we develop a neural network model to evaluate call quality degradations in two separate categories: echo and degradations from other sources. We show that our model is accurate as measured by correlation with human subjective quality ratings. Our tool can be used effectively to stack rank echo cancellation models. AECMOS is being made publicly available as an Azure service.

\end{abstract}
\section{Introduction}
\label{sec:intro}

Acoustic echo arises when a near end microphone picks up the near end loudspeaker signal and a far end user hears their own voice. The presence of acoustic echo is a top complaint in user ratings of audio call quality.

\emph{Acoustic echo cancellers} (AECs) significantly improve audio call quality by canceling echo. This is achieved by first comparing the transmitted signal to the received signal, processing for delay and signal distortions, and then subtracting to remove the echo component. The goal is to only transmit a clean near end signal.

Recently, there has been a lot of innovation in AECs. The rise of deep learning has led to better performing models as compared to their classical counterparts \cite{Seidel2021Y2NetFF, pfeifenberger21_interspeech, 9053508, 2021amazon, 9054541}. Also, hybrid models, combining both classical and deep learning methods such as using adaptive filters and \emph{recurrent neural networks} (RNNs) \cite{peng21f_interspeech, ma2020acoustic}, have shown great results \cite{sridhar2021icassp, Zhang2021FTLSTMBC, cutler2021interspeech}. With the rapid advancements in the field of echo cancellation, it becomes ever more critical to be able to objectively measure and compare the performance of AECs.

We propose a speech quality assessment metric for evaluating echo impairment that overcomes the drawbacks of conventional methods. Our model, called AECMOS, directly predicts human subjective ratings for call echo impairment. It can be used to evaluate the end-to-end performance of AECs and to rank different AEC methods based on (degradation) \emph{mean opinion score} (MOS) estimates with great accuracy. 

Our model architecture is a deep neural network comprising of convolutional layers, GRU (gated recurrent unit) layers, and dense layers. AECMOS is trained using the ground truth human
ratings obtained following guidance from ITU-T Rec. P.831 \cite{itut_p831}, ITU-T Rec. P.832 \cite{itut_p832} and ITU-T Rec. P.808 \cite{itut_p808} as described in  \cite{cutler2020crowdsourcing}.

AECMOS is extremely effective in aiding with the development and research of AECs. We are providing the AECMOS as an Azure service for other researchers to use as well. The details of the API are at \url{https://github.com/microsoft/AEC-Challenge/tree/main/AECMOS}. We have already received over $50$ requests for accessing our service ranging from university researchers to big companies. For examples of AECMOS (aka DECMOS) use in AEC model development we refer the reader to \cite{peng21f_interspeech, Seidel2021Y2NetFF} which include winners of the INTERSPEECH 2021 Acoustic Echo Cancellation Challenge \cite{cutler2021interspeech}.

AECMOS does not require a clean speech reference for the near or far end, nor a quiet environment. A clean reference is typically not available in non-artificial scenarios, and the test set that we use for evaluating AECMOS can be much more realistic and representative than artificial scenarios. Our test set was selected from over 5000 different scenarios, each with different room acoustics, different devices, and different human speakers. AECMOS can be used in actual customer calls to monitor the quality of real calls; it is not restricted to lab or development usage, but has operational utility.

\section{Related work}

Common methods of evaluating AEC models \cite{2018zhang, 2019fazel, 2003guerin, santos2018speech} include using intrusive objective measures such as  \emph{echo return loss enhancement} (ERLE) \cite{g168} and  \emph{perceptual evaluation of speech quality} (PESQ) \cite{p862}.  ERLE can only be measured when having access to both the echo and processed echo signals, in a quiet environment without near end speech. ERLE can be approximately measured by: $\displaystyle
ERLE \approx 10\log_{10} \frac{\mathbb{E}[y^2(n)]}{\mathbb{E}[e^2(n)]} 
$
where $y(n)$ is the microphone recording of the far end signal (with no echo suppression), and $e(n)$ is the residual echo after cancellation. PESQ requires a clean speech reference in addition to the degraded speech.

Unfortunately, metrics such as ERLE and PESQ are often not well correlated with human subjective ratings of call echo degradation quality \cite{cutler2020crowdsourcing}. This is especially true in the presence of background noise or double talk \cite{Avila2019, cutler2020crowdsourcing}.   
Carrying out laboratory tests following standards such as ITU-T Rec. P.831 \cite{itut_p831}, while more accurate, is expensive, time consuming, and not a scalable solution. 

There are objective standards to help characterize AEC performance. IEEE 1329 \cite{ieee1329} defines metrics like terminal coupling loss for single talk (TCLwst) and double talk (TCLwdt), which are measured in anechoic chambers. 
TIA 920 \cite{tia920} uses many of these metrics but defines required criteria. 
ITU-T Rec. G.122 \cite{g122} defines AEC stability metrics, and ITU-T Rec. G.131 \cite{ITU-G131} provides a useful relationship of acceptable Talker Echo Loudness Rating versus one way delay time.
ITU-T Rec. G.168 \cite{g168} provides a comprehensive set of AEC metrics and criteria. However, it is not clear how to combine these dozens of metrics to a single metric, or how well these metrics correlate to subjective quality. 

Commercially available objective metrics include EQUEST \cite{equest} which measures single talk echo performance. 
An objective metric for double talk echo performance is ACOPT 32 \cite{acopt32} which implements the 3GPP standard TS 26.132 \cite{3gpp}. ACOPT 32 is of limited use in real call scenarios as it requires the same near-end signal to be played twice once as single talk and once as part of double talk. Furthermore, it is not straightforward to interpret the model outputs.  

\section{Data}
\label{sec:data}

We use supervised learning to train AECMOS. Each example in the dataset consists of three audio signals: near end microphone signal, far end signal, and the output from an echo canceller which we call the enhanced signal. The label for a given example is its (degradation) Mean Opinion Score (MOS) obtained from crowdsourcing as described in \cite{cutler2020crowdsourcing}.

In generating the data set, we distinguished between single talk and double talk scenarios. For the near end single talk case, we asked for the overall quality \cite{itut_p808}. For far end single talk, we asked for echo ratings \cite{itut_p831}. For double talk, we asked for ratings for both echo annoyance and other degradations in two separate questions \footnote{Question 1: How would you judge the degradation from the echo?
Question 2: How would you judge other degradations (noise, missing audio, distortions, cut-outs)?}. All impairments were rated on the degradation category scale (from 1:Very annoying to 5: Imperceptible). The ratings were then used to obtain a MOS label for the examples. For near end single talk, the echo label was set to $5$ and for far end single talk the degradation label was set to $5$. We found including ratings for double talk non-echo degradations to improve the correlation between experts and naive raters \cite{cutler2020crowdsourcing}. 

The model was trained on $64,013$ samples ranging in duration from $3$ seconds to $14.5$ seconds with a mean of $8.2$ seconds, for a total of 145.8 hours of data. The training data included 17 submitted models with a total of 14K audio clips from the ICASSP 2021 AEC  Challenge \cite{sridhar2021icassp}. The breakdown of the training data set by scenario is: $45.6 \%$ near end single talk, $26.7 \%$ far end single talk, and $27.7\%$ double talk.

\begin{figure}[h!]
\centering
\includegraphics[scale = 0.54]{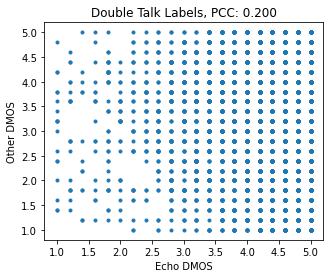}
\caption{Test set double talk ground truth labels.}
\label{fig: dt_ground}
\end{figure}

The test set consists of submissions from the INTERSPEECH 2021 AEC Challenge \cite{challenge, cutler2021interspeech}. There were a total of $14$ contestants submitting their work for $300$ double talk, $300$ far end single talk, and $200$ near end single talk examples. In addition, we included the enhanced signals of $4$ of our own deep models along with $4$ digital signal processing based models.  In this manner, we obtained a total of  $17,600$ enhanced speech signals or $55$ hours of data. The ground truth labels were gathered via crowdsourcing using the tool \cite{cutler2020crowdsourcing}, where we collected $5$ votes per clip. \autoref{fig: dt_ground} shows that, in the case of double talk, a wide variety of echo and other degradation MOS score behaviours co-occur. 

\section{Model}
\label{sec:model}

Advancements in deep learning have shown great potential in various speech enhancement tasks \cite{2021amazon, santos2018speech}, including for speech quality assessment \cite{reddy2021dnsmos, 2021dolby}. Here we propose a deep neural network to replace humans as call quality raters for speech impairment due to echo and other degradations.

\subsection{Features}
\label{sec:features}
The AECMOS model takes as input three signals: near end microphone signal, far end signal, and the output from an echo canceller also referred to as the enhanced signal. The task is to evaluate the quality of the enhanced signal with regard to echo impairment on a human subjective rating scale. Observe that the very nature of acoustic echo necessitates a comparison of signals. Knowledge of both the near end microphone signal and the far end signal is needed to determine whether the enhanced signal contains echo as opposed to some background noise, for example. This is in contrast to evaluating noise suppression quality which can be done successfully without reference signals \cite{reddy2021dnsmos}.

In addition to the three input signals, AECMOS also takes an optional scenario marker as part of the input. This marker encodes which of the three scenarios we are in: near end single talk, far end single talk, or double talk. For online deployment, the marker is not used. For offline AEC model evaluations, when scenario information is readily available, activating the scenario marker improves model performance as shown in \autoref{table: ablate1}. In our training set, each sample has an associated scenario label that was created at the time of the generation of the training data.
We found it effective to prepend a one-hot vector of a fixed length to the three model input signals indicating which signal(s) should be considered active for a particular sample.

We used {\it micro augmentations} during training to increase model stability. By this, we mean imperceptible data augmentations: removing the initial 10 ms of the near end microphone signal, or changing the energy level by 0.5 dB.

Details about the data set and the crowdsourcing approach for obtaining the ground truth labels appear in Section \ref{sec:data} and \cite{cutler2020crowdsourcing, challenge, reddy2021dnsmos}.

\subsection{Architecture}

For the model architecture, we explored different configurations of convolutional models. A significant boost in MOS prediction correlations with the ground truth labels came from incorporating GRU layers into the model. \autoref{table: 1} shows the architecture for the best performing model. In developing the model, we saw that for all models the double talk scenario posed the hardest challenge. As shown in \autoref{table: ablate1}, incorporating GRU layers improved model performance most significantly for double talk.

The input to the model is a stack of three log power spectrograms obtained from the near end, far end, and enhanced signals. The spectrograms were computed with a DFT size of 512 and a hop size of 256 over clips that were sampled at 16 kHz. Finally, we calculate the logarithm of the power. For an 8 second clip, this results in an input dimension of $541 \times 257$. AECMOS handles variable length inputs natively. 
\begin{table}[ht]
\centering
\caption{AECMOS architecture} 
\label{table: 1}
\begin{tabular}{l r } 
\hline\hline 
Layer & Output Dimensions  \\ [0.5ex] 
\hline 
Input: $3 \times 541  \times  257$  &\\
\hline 
Conv: 32, \; $(3 \times 3)$, LeakyReLU & $(32, 270, 128)$  \\ 
MaxPool: (2 x 2), Dropout(0.4)  \\
\hline 
Conv: 64, \; $(3 \times 3)$, LeakyReLU & $(64, 135, 64)$  \\ 
MaxPool: (2 x 2), Dropout(0.4)  \\
\hline
Conv: 64, \; $(3 \times 3)$, LeakyReLU & $(64, 67, 32)$  \\ 
MaxPool: (2 x 2), Dropout(0.4)  \\
\hline
Conv: 128, \; $(3 \times 3)$, LeakyReLU & $(128, 33, 16)$  \\ 
MaxPool: (2 x 2), Dropout(0.4)  \\
\hline
Global MaxPool  & (1, 128)  \\ 
\hline
Bidirectional GRU: 128, NumLayers 2 \\
HiddenUnits 64, Drouput(0.2)   & $(1, 128)$  \\ 
\hline
Dense: 64, \; LeakyReLU \;  Dropout(0.4) & $(1, 64)$  \\ 
\hline
Dense: 64, \; LeakyReLU \;  Dropout(0.4) & $(1, 64)$  \\  
\hline
Dense:  2, \; 1 + 4* sigmoid  & $(1, 2)$  \\ 
\hline 
\end{tabular}
\end{table}

The model was trained on $8$ GPUs with a batch size of 10 per GPU. We used the Adam optimizer \cite{adam} and MSE loss function until the loss saturated. 

The model outputs two MOS predictions on a scale of $1-5$: one for echo and one for other MOS. These outputs correspond to the ground truth labels as described in Section \ref{sec:data}.

\section{Experiments}
\label{sec:experiments}

We evaluate the accuracy of our model by measuring the correlation between the predictions of our AECMOS and the ground truth human ratings. 

More precisely, we used the enhanced signals from a total of 22 models. For each model, the outputs were produced for 300 double talk, 300 far end single talk, and 200 near end single talk examples. See Section \ref{sec:data} for more details. For each audio clip, we obtained the objective MOS rating using our model and the crowdsourced MOS score. 

\subsection{Results}
\label{sec:results}
In order to evaluate our model, we calculate the PCC for the AECMOS predictions versus the ground truth MOS labels. 

For each far end single talk example, we also calculate the ERLE, PESQ, and EQUEST score.
\begin{figure}[h!]
\centering
\includegraphics[width = 7cm]{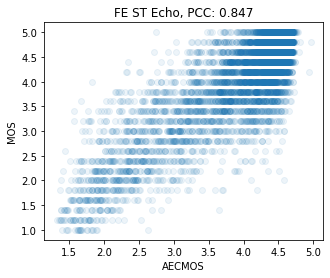}
\caption{far end Single Talk Per Clip: AECMOS}
\label{fig: fest_echo}
\end{figure}
As seen in Table \ref{table: corrs1}, in the far end single talk scenario, AECMOS outperforms ERLE, PESQ, and EQUEST (evaluated with 150 wideband test conditions). 

For the near end single talk scenario, we can compare our AECMOS with DNSMOS model which was developed for evaluating noise suppression models \cite{reddy2021dnsmos}.
%
%
%
AECMOS has a more difficult task than DNSMOS: evaluate both echo and other degradations and do so independently of each other. Nonetheless, we believe that AECMOS has very good potential to improve in the near end single talk category. For one, DNSMOS was trained on about $120,000$ audio clips \cite{reddy2021dnsmos}, while AECMOS saw only about half that many for training and only a quarter as many, roughly $30,000$, were near end single talk clips. With this in mind, AECMOS performance is very promising.

%
\begin{table}[h!]
\caption{Per Clip PCC for AECMOS, and other commonly used metrics: DNSMOS, ERLE, PESQ, EQUEST.} 
\label{table: corrs1}
\centering
\begin{tabular}{p{2.3cm}p{.6cm}p{.6cm}p{.6cm}p{.6cm}p{1.1cm}}
\hline 
\hline
& AEC MOS & DNS MOS & ERLE & PESQ & EQUEST\\ 
\hline
ST far end DMOS & 0.847 && 0.541 & 0.710 & 0.686 \\
ST near-end MOS & 0.611 & 0.640 \\
DT Echo DMOS & 0.582 &\\
DT Other DMOS & 0.751
\end{tabular}
\end{table}

Expectedly, the most challenging scenario to evaluate is the double talk scenario. Here the model needs to evaluate separate qualities, echo and other degradations, simultaneously yet independently of each other. Measuring and improving double talk performance is important as not being able to interrupt others speaking has been shown to impair meeting inclusiveness and participation rate \cite{cutler2021meeting}.

For evaluating the stack ranking of different echo cancellers, we compute the average of ratings across the entire test set for each model. We calculate the same for AECMOS ratings. Finally, we calculate the Spearman's Rank Correlation Coefficient (SRCC) between the two. The results are given in \autoref{table: corrs2}. We report an SRCC of $0.969$ and a PCC of $0.996$ in the far end single talk scenario, which is the most common scenario for echo cancellation. We note that the best performing submitted models were very close to each other in the contest. In fact, so much so that the human MOS rankings were within error bars of each other \cite{cutler2021interspeech}. 

\begin{table}[h!]
\centering
\caption{AECMOS Per Contestant PCC and SRCC: All Scenarios refers to far end single talk and double talk for echo; and near end single talk MOS and double talk Other MOS.}
\label{table: corrs2}
\begin{tabular}{p{3.5cm}p{.6cm}p{0.6cm}}
\hline
\hline
& PCC & SRCC \\ 
\hline
All Scenarios Echo DMOS & 0.981 & 0.970 \\
All Scenarios (Other) MOS &0.902 & 0.954 \\
ST far end Echo DMOS & 0.996 & 0.969\\
ST near end MOS & 0.923 & 0.831\\
DT Echo DMOS & 0.898 & 0.863 \\
DT Other DMOS&  0.927 & 0.955 \\
\end{tabular}
\end{table}

\begin{figure}[h!]
\centering
\includegraphics[width = 7cm]{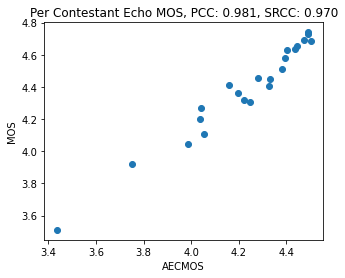}
\label{fig: cont_echo}
%
\centering
\includegraphics[width = 7cm]{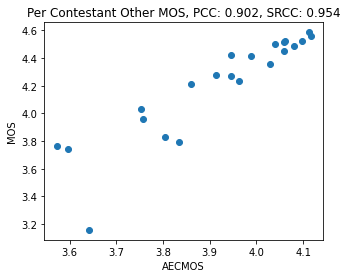}
\caption{Per contestant: echo degradation (far end single talk and double talk) and other MOS (near end single talk and double talk).}
\label{fig: cont_deg}
\end{figure}

\section{Ablation Study}

We experimented with various architectures, features, and training methods. In this section, we give an overview of the key findings.

We started out with a baseline model consisting of $5$ convolutional layers followed by $3$ dense layers. The input to all models was a stack of three features: log power of STFT applied to the near end, far end, and enhanced signal. All models were trained using micro augmentations as described in Section \ref{sec:model}.

The first improvement in our model's performance came from including scenario labels as part of the model input. As we were exploring avenues for model development, we conducted a small experiment where we incorporated label information in the loss function and asked the model to predict scenario labels in addition to the MOS labels. Curiously, the model predicted the far end single talk scenario $87\%$ of the time while only $52\%$ of the labels were actually far end single talk with the remaining labels being double talk. Introducing an optional scenario marker for offline use helped model performance as described in Table \ref{table: ablate1}. More discussion of input features can be found in Section \ref{sec:features}.

The second improvement came about when we were investigating how to improve double talk performance. Experiments with convolution kernel size led us to believe that our model was having difficulties incorporating information along the temporal axis. Introducing GRU layers into the model remedied this issue. In adding a new layer, we needed to remove a convolutional layer so that we would not be down sampling too aggressively before reaching the GRU layer. Table \ref{table: ablate1} summarizes the aforementioned key improvements.

\begin{table}[h!]
\centering
\caption{Per Clip Pearson Correlation Coefficients: Baseline Convolutional Model; add scenario markers to model input; remove a convolution layer and add a GRU layer to obtain AECMOS.} 
\label{table: ablate1}
\begin{tabular}{p{3.45cm}|p{1cm}|p{1.4cm}|p{.9cm}}
\hline \hline
& Baseline   &  + scenario & + GRU  \\
\hline
All Scenarios Echo DMOS & 0.732 & 0.746 & 0.797 \\
All Scenarios (Other) MOS & 0.735 & 0.775 & 0.802 \\
ST far end Echo DMOS    &  0.780  &  0.825 & 0.847 \\
ST near end MOS & 0.434 & 0.534 & 0.611 \\
DT Echo DMOS & 0.458   & 0.422  & 0.582 \\
DT Other DMOS & 0.577  & 0.657 & 0.751\\
\hline
\end{tabular}
\end{table}

We also experimented with using log power of Mel spectrogram for model input features. Mel spectrogram corresponds well with human subjective hearing and has been successfully used in evaluating noise suppression \cite{reddy2021dnsmos}. Our best model that takes Mel spectrogram features as input uses 160 Mel bins. While we experimented with different settings, we found a consistent behavior that the Mel models achieve lower correlation scores for echo and higher scores for other degradations. This matches our intuition as detecting the presence of echo is less dependent on human subjective experience than classifying a sound as noise. 

\begin{table}[h!]
\centering
\caption{Per Clip Pearson Correlation Coefficients: AECMOS; AECMOS trained with Mel spectrogram features.} 
\label{table: ablate2}
\begin{tabular}{ p{3.5cm}| p{1.5cm} |p{1.9cm}}
\hline \hline
& AECMOS & AECMOS Mel  \\ 
\hline
All Scenarios Echo DMOS & 0.797  &  0.742 \\
All Scenarios (Other) MOS & 0.802  & 0.819 \\
ST far end Echo DMOS  & 0.847 & 0.739 \\
ST near end MOS & 0.611 & 0.604 \\
DT Echo DMOS & 0.582 & 0.553 \\
DT Other DMOS & 0.751 & 0.772 \\
\hline
\end{tabular}
\end{table}

Finally, we experimented with the self-teaching paradigm \cite{reddy2021dnsmos} and training with bias-aware \cite{mittag2021bias} loss. Interestingly, neither provided significant improvements in PCC.

\section{Improvements due to growing training data.}

We conjecture that our model's performance improves with more training data. In particular, we have observed that increasing the size of our training data from 64K audio clips to 108K clips and testing on 5K clips, gives a model that is just as accurate for inference whether or not it uses scenario label information. The difference in correlations was roughly $1\%$ across all scenarios. However, since we used more data for training we could not validate this claim as thoroughly as we have done for our work in the previous sections of this paper. 

\section{Conclusions}
\label{sec:end}

Our AECMOS model provides a speech quality assessment metric that is accurate, expedient, and scalable. It can be used to stack rank echo cancellers with very good accuracy and thereby accelerate research in echo cancellation. In the future, we would like to further improve the model by exploring additional data augmentations and learning custom filter banks.


\bibliographystyle{IEEEbib}
\bibliography{strings,refs}

\end{document}